\documentstyle[aps,preprint,eqsecnum]{revtex} \baselineskip 10pt
\def\matrix#1#2#3#4{\left(\begin{array}{cc}#1 & #2 \\ #3 & #4 \end{array}\right)}

\def\eqr#1#2#3{\begin{eqnarray}#1\nonumber\\#2\label#3}
\begin{document}
\preprint{\today}
\title{D-BRANE INTERACTION IN THE TYPE IIB MATRIX MODEL}


\author{Bhabani Prasad Mandal and Subir Mukhopadhyay}

\address{
Institute of Physics, Sachivalaya Marg,\\ Bhubaneswar-751005, India,\\
\mbox{email: bpm, subir@iop.ren.nic.in } \\}

\vspace{.4in}

\maketitle


\begin{abstract}
We calculate the potential  between  bound states of D-branes of
different dimension in IIB matrix model upto one loop order and
find nice agreement with the open string calculations in short and large
distance
limit. We also consider the scattering of bound states of D-branes, 
calculate the scattering phase shift and analyze the effective potential
in different limits. 
\end{abstract}

\newpage


\section{Introduction}

Recently, a non-perturbative formulation (matrix theory) \cite{bfss} has been 
proposed for the M-theory at
infinite momentum frame. 
Making use
of the fact, that
at infinite momentum frame
all other degrees of freedom with finite momentum get decoupled, 
the M-theory
has been shown to be described by that of a system of large number of 
D-particles. The theory is given by a large-$N$ super Yang-Mills (SYM) theory in $(0+1)$
dimension. A number of issues have been tested such as supergraviton spectrum,
D-brane interaction and various kinds of compactifications \cite{bfss,raj,ram,sei,bera,lyfm,lyf}. 

This model has inspired another large-$N$ super Yang-Mills theory in $(0+0)$ 
dimension \cite{ikkt} 
which
is proposed to describe the non
perturbative physics of type IIB string. 
It has its origin in the fact that a large-$N$ reduced SYM theory describes string at a
proper double scaling limit \cite{bala}. 
However, a closer look 
reveals that
this is the theory of D-instanton (apart from a chemical potential term in the 
action) and one can use a similar infinite momentum frame argument to arrive at
this theory \cite{tyl}. This theory, by compactification on one circle and a 
subsequent wick rotation leads to the BFSS matrix model \cite{bfss}. The massless 
spectrum and the
light-cone string theory can be recovered from this model \cite{ikkt,bala,mli}.    

In this paper, we shall study the interaction of the D-branes in the latter formulation 
which, among other 
things, serves as a consistency check for the type IIB matrix theory and helps
to understand the relation between these two matrix theories. 

The D-branes are the BPS states of the string theory which couple to the RR fields.
Usually a p D-brane couple to a (p+1) form potential. By virtue of the interaction
term $\int C \wedge tr(e^{(iF)})$, a p-brane with electro-magnetic field turned on in
its world volume can interact with a brane of lower dimension and thus can form  
a non-threshold bound states \cite{lyf}. 
These composite objects with large electric or magnetic field 
turned on in its world volume, {\it i.e.} a D-brane coupled to a large number of 
lower dimensional D-branes, come out naturally in the matrix 
theory as the classical solutions of the equation of motion. So from Matrix
theory one can study the interactions between these D-branes.
Since the Matrix theory describes the interaction mediated through the open string,
it should reproduce the short distance behavior. Further, due to the presence of 
a large number of lower dimensional D-branes it should also reproduce the long 
distance limit.

In the BFSS theory \cite{bfss}, the interaction among D-branes has been 
studied in one loop order of the matrix theory leading to a satisfactory 
agreement with 
the open string calculation upto one loop both at short and long distance limit
\cite{bera,lyfm,lyf}. In the IKKT theory \cite{ikkt}, the long distance limit of the interaction
potential for some configurations breaking half of the SUSY, has been
compared with the result obtained from the Born-Infeld action and nice agreement has been reached \cite{che1}.
The short distance limit has also been analyzed in \cite{mak,che} but the lower
dimensional branes have not been considered. 

Here we shall study the potential and the scattering phase
shift from the IIB matrix theory and compare the results with those obtained 
from the open and the closed string calculations. 

The plan of this paper is as follows. In the next section we will discuss some
preliminaries concerning the IKKT model. Then in the two following sections the 
potential energy and the scattering between two D-branes of odd dimensions will
be discussed. Finally we will conclude discussing our results and possible
further extensions.

\section{Preliminary}

In this section, we describe the necessary preliminaries required for the study
of interaction among D-branes in the type IIB matrix model \cite{ikkt}.  Let us start with the action
which is obtained by reduction of ten dimensional Euclidean SYM 
theory with $SU(N)$ gauge group reduced to zero dimension. 
\begin{equation}
S= \alpha \{-\frac{ 1}{4} Tr [A_\mu ,A_\nu]^2 - \frac{ 1}{2} Tr (\bar{\psi}
\Gamma ^\mu [A_\mu,\psi])\} +\beta N
\label{action}
\end{equation}
where $A_\mu$ and $\psi$  are gauge fields in the $SU(N)$ adjoint and so we 
can consider them to be $N\times N$ Hermitian matrices.
The parameter $N$ is
considered
to be a dynamical variable, $\alpha$ and $\beta$ are related to the string coupling.
The third term implies that the sum over $N$ in partition function should be
taken with the weight factor $e^{-\beta N}$. Later the sum over $N$ is
replaced by a double scaling limit \cite{bala}.

This action can be viewed either as obtained from type IIB theory by taking the Schield
gauge \cite{ikkt} or as the action for $N$ D-instantons \cite{tyl} apart from the last term. It 
has the manifest Lorentz
invariance and $N=2$ supersymmetry. The SUSY transformations are given by
\begin{eqnarray}
\delta^{(1)}\psi &=& \frac{i}{2} [A_\mu, A_\nu]\Gamma^{\mu\nu}\epsilon ,~~~ 
\delta^{(2)}\psi = \xi, \nonumber \\ 
\delta^{(1)}A_\mu &=& \bar\epsilon\Gamma_\mu\psi, ~~~ \delta^{(2)}A_\mu = 0.
\label{susy}\end{eqnarray}

The equation of motion is given by
\begin{equation}
[A^\mu,[A_\mu,A_\nu ]] = 0\;\;\;\;\; [ A_\mu, (\Gamma ^\mu \psi)_ \alpha ] =0
\label{eom} 
\end{equation}
The solution of the equation of motion correspond to non-perturbative states in
IIB theory. In particular, those cases for which $[A_\mu,A_\nu] = {\rm
c-number}$, are interesting as they preserve half of the supersymmetry
and are interpreted as BPS states. The odd p-branes, being BPS states in
type IIB theory, are associated with this type of solutions.

So a plausible choice for the p-D-brane solution is 
\begin{eqnarray}
A_\mu^{cl} &=& B_\mu,~~~~ {\mathrm for}~~ 0\leq\mu\leq p ,~~~~\psi =0 \nonumber \\
A_\mu^{cl} &=& 0 ~~~~{\mathrm otherwise} 
\label{brane} \end{eqnarray}
where $B_\mu$'s are the $N\times N$ Hermitian matrices satisfying 
$[B_a,B_b] =-ic_{ab}I$, and $ a,b =0,\cdots p$ with $c_{ab}$ is a c number. 

A consistent supersymmetry algebra is yet to be constructed (an attempt was 
made in \cite{che}) but the central charge of a p-brane can be taken as \cite{sei,mak,che} 
\begin{equation}
Z_{i_1,\cdots i_p} = [A_0,A_{i_1}][A_{i_2},A_{i_3}]\cdots[A_{i_{p-1}},A_{i_p}]
\end{equation} which is consistent with this solution. But this means that the p-branes 
contain non zero electric or magnetic flux and so it actually correspond to
a large number of lower dimensional branes coupled to it.

In general, this solutions represent D-branes
of infinite extension. However, we would like to deal with finite ones for
the time being. So we wrap the p-brane on a torus of dimensions $L_i$. 
For the sake of convenience of the calculation, let us make use of the  Lorentz 
transformation
to cast the $c_{ab}$ in a block diagonal form with eigen values $\omega_i$'s . 
Then the
eigenvalues, related to $L_i$ by the relation
\begin{equation}
2\pi \omega _i =\frac{  L_{2i-2}L_{2i-1}}{N_i}
 \label{omega}
\end{equation}
represents the amount of flux linked with the respective planes. 

At the end of the calculation we will take the infinite limit for the $L_i$'s
and $N_i$ keeping $\omega_i$ finite. 
So we can consider the $B_\mu$'s to be Hermitian operator
on some Hilbert space and the matrix index can be taken to be continuous.
Further, we, use the Schr\"odinger representation for the Hilbert space operators
$B_i$ and write
\begin{equation}
B_{2i-2} =i \omega _i \partial _i ;\;\;\; B_{2i-1} =q_i \ \ \ i,j =1,2\cdots
\label{B_i}
\end{equation}
where $\partial_i\equiv \partial/\partial q_i$ and $q_i$ represent conjugate
operators on the Hilbert space. 
Finite values of $L_i$'s imply that the
eigenvalues of $q_i$ are distributed in $ [- \frac{ L_q}{2}, \frac{ L_q}{2}]$.

We will consider the interaction between p-branes upto one loop. Summing over
the fluctuations and considering only the quadratic terms, the one
loop effective action  of the matrix model can be written in a compact form
\cite{ikkt}
\begin{equation}
W= \frac{ 1}{2} Tr \ln (P^2 \delta_{\mu\nu} -2iF_{\mu\nu} -\frac{ 1}{4}
Tr \; \ln (P^2+ \frac{ i}{2} F_{\mu\nu} \Gamma ^{\mu\nu}) 
( \frac{ 1+ \Gamma _{11}}{2})- Tr \ln (P^2)
\label{W0}
\end{equation}
where $P_\mu $ and $F_{\mu\nu}$ are operators acting on the space of Hermitian 
matrices and are
defined by 
\begin{equation}
P_\mu X = [A_\mu^{cl},X],\;\;\;\;\; F_{\mu\nu} X = i[[A^{cl}_\mu,A_\nu^{cl}],X]
\label{eff0}
\end{equation}
In other words $P_\mu$ and $F_{\mu\nu}$ are the adjoint representation of 
$A_\mu$ and $F_{\mu\nu}$ respectively. 

Here we will restrict ourselves to the one loop order. We will get the classical
solution corresponding to the configuration and obtain the effective action from
(\ref{W0}). This analysis can be extended to higher loop orders but that will
involve \cite{bec}
explicit Feynman diagrams and the effective action will be much 
more complicated. 

\section{ Potential between branes}

In this section we will consider the potential between two branes of dimensions
$p$ and $q$ respectively. A configuration of two parallel branes of same dimension is known to 
saturate the BPS limit and so the potential is zero. For $p\not=q$ in the
generic case it breaks the supersymmetry and so the potential is nonzero.
All these have been investigated
by usual D-brane analysis \cite{lyf1,tasi,bac}. However, in matrix model, instead of 
getting pure D-branes , the classical solutions represent D-branes with electric
and magnetic fluxes turned on in its world volume, which, as stated earlier,
represent configurations of D-branes of different dimensions \cite{ram,lyfm,lyf} . Our aim is to 
carry out similar kind of investigation in
the IIB matrix model framework.  

We shall consider the two brane configuration as the classical solution
of the equations of motion (\ref{eom}). Then we will sum over all fluctuations 
around it upto one loop order using (\ref{eff0}) and get the interaction
potential as described in Ref. \cite{ikkt}.

Let us start with the two branes of $p$ and $q$ dimensions parallel to the
$X^{1\cdots p}$ and $X^{1\cdots q}$ planes 
and place them at a distance $\pm b/2$ along $X^{q+1}$ respectively. 

The classical configuration is given by,
\begin{eqnarray}
A_i &=&\left ( \begin{array}{cc} B_i&0\\0&B_i \end{array} \right );\ \ 
{\mathrm for} \ \ 0\le i\le p; \ \ A_i 
= \left (  \begin{array} {cc}
B_i&0\\0&0
\end{array} \right );\ \ {\mathrm for} \ \ p+1\le i\le q; \nonumber \\
A_{q+1} &=& \frac{b}{2} \left( \begin{array} {cc}
1&0\\0&{-1} \end{array} \right ),\ \ A_i=0 \ \ {\mathrm for} \ \  
i>q+1,     \label{sol1}
\end{eqnarray}
where $B_i$'s are the Hermitian operators on Hilbert spaces mentioned earlier.

So the B's in the Scro\"dinger representation are given by
$B_{2i-2}=i\omega\partial_i$ , $B_{2i-1}=q_i$ for $1\le i\le l$
with the value of $\omega$ is given by (\ref{omega}).

In order to get the potential we have to sum over the fluctuations using (\ref{W0})
around this classical solution (\ref{sol1}). The first thing we need is the 
adjoint representation of the operators corresponding to $A_\mu$ given by the
action on an arbitrary Hermitian matrix according to the (\ref{eff0}). This is
given by 
\begin{eqnarray}
P_{2i-2}\left(\begin{array} {ll} X&Y\\ Y^\dagger &Z \end{array} \right) &=& i \omega _1( \partial
_i^1+\partial _i^2 ) \left( \begin{array} {ll} X&Y \\ Y^ \dagger & Z
\end{array} \right)
;\ \ 
P_{2i-1}\left(\begin{array} {cc} X&Y\\ Y^\dagger &Z \end{array} \right) 
= (q_i^1 - q_i^2)
\left(\begin{array} {cc} X&Y\\ Y^\dagger &Z \end{array} \right) 
\nonumber \\ 
P_{q+1}
\left(\begin{array} {ll} X&Y\\ Y^\dagger &Z \end{array} \right) 
&=& b/2
\left(\begin{array} {ll} 0&Y\\ Y^\dagger &Z \end{array} \right) 
\end{eqnarray}
where each entry is an infinite dimensional matrix {\em e.g.} $X=X(q^1,q^2)$
and also $P_1 = i\omega_1\partial_1\delta(q^1_1 -q^2-1)$, 
$q_1 = q^1\delta(q^1_1-q^2_1)$ etc. are infinite dimensional matrices.

The adjoint representation of the $F_{\mu\nu}$ can be obtained by using
(\ref{eff0}) from which we see that the only $F_{i,i+1}$ components have
non-trivial actions. Further they act only on the matrices of the form 
$ \left(\begin{array} {cc} 0&Y\\ Y^\dagger &0\end{array} \right)$
which correspond to a non-zero eigenvalue. The eigenvalues are
$\omega_i$ with $k<i\le l$.

Similarly the eigenvalues of $P^2$ can be obtained as
\begin{equation}
E = 2\sum_{i=1}^k (p_i^2+\tilde q^2_i) +\sum_{k+1}^l (p_i^2+q_i^2)
\label{E}
\end{equation}
where $[p_i, q_j] = -i\omega_i\delta_{ij}$ and other commutators are 
vanishing. It looks like Hamiltonian of $k$ free particles and $l-k$
oscillators.

Collecting all these three we get the interaction
potential from the (\ref{W0}). After Simplification it becomes 
\begin{equation}
W= \frac{ 1}{2} \sum_1^\Delta Tr\ln (E-2a_i) + Tr\ln (E+2a_i)-2^{2-\Delta}
\sum_{\{\lambda _i\}} Tr\ln (a_i \lambda _i) + (4-\Delta)\ln E
\label{ eff2}
\end{equation}
where $\lambda_i=\pm1$ , $\Delta=l-k$ and $a_i = \omega_{k+i}$.

In order to compare this with the existing result it is better to write the
potential in the form of a proper time integral. Using the identity \cite{mak},
\begin{equation}
\ln a = -\int \!\frac{ds}{s}\exp{ (-sa)}
\label{ln}
\end{equation}
we can write the potential as 
\begin{equation}
W= \int \frac{ds}{s} Tr e^{-sE} \left \{ \sum_1^\Delta\cosh(2s\omega_i) -4
\prod_{i=1}^\Delta\cosh (s\omega_i) +(4-\Delta) \right \},
\label{w5}
\end{equation}
where $E$ is given by (\ref{E}).

The trace can be evaluated in a straightforward manner and is given by
\begin{equation}
Tr e^{-sE} = \frac{ N^{2k}}{L_0\cdots L_p}(\frac{ \pi}{2s})^k
\exp{-b^2s}\prod_{i=k+1}^l(2\sinh \omega _is)^{-1}
\label{ }
\end{equation}

Using this we can get
\begin{equation}
W= C_p \prod_{i=1}^k \frac{ 1}{|2\pi \omega _i|^2}\int \frac{ ds}{s}
\exp(-b^2s)(\frac{ \pi}{2s}^k. \frac{
\sum_{i=k+1}^l \cosh(2s\omega_i)-4\prod_{k+1}^l\cosh(s\omega_i) +4
-\Delta}{\prod_{i=k+1}^l 2 \sinh (\omega _is) }
\label{w} 
\end{equation}
where $C_p$ is the volume of the hyperplane which is common to 
both the branes.

Note that this expression is almost identical to the expression of the phase
shift obtained in the scattering of D-particle and 4-branes in \cite{lyf}.
Only difference is that they consider the scattering while here we are 
considering the static potential. This confirms the T-duality between the
two matrix models.
 
As mentioned earlier this represents the potential between $p$ and $q$-branes
with stack of lower dimensional branes attached to these. 
In order to compare this with the result obtained from open string calculation
upto one loop we consider the specific brane configurations.

For the sake of brevity let us consider the 3-brane and 1-brane configuration. So
we put $q=3$ and $p=1$. The 3-brane has a magnetic field turned on in its
world volume on the $X^{23}$ plane with strength 
$\int_{C_{23}} F \equiv \omega_2$. Besides, both the 3 and the 1-brane has
an electric field in their world volume with $\int_{C_{01}} F \equiv \omega_1$.
So the 3-brane, considered in this matrix model actually represents a bound state of
a 3-brane, $\frac{L_2 L_3}{2\pi\omega_2}$ D-strings along the $X^1$ and 
$\frac{L_0 L_1}{2\pi\omega_1}$ D-instantons and 
similarly the 1-brane represents a bound state of a 1-brane with
$\frac{L_0 L_1}{2\pi\omega_1}$ D-instantons. 

In this case the potential is reduced to
\begin{equation}
W= L_1 \frac{1}{|2\pi \omega _1|^2}\int \frac{ ds}{s}
\exp(-b^2s)(\frac{ \pi}{2s}). \frac{
\cosh(2s\omega_2)-4\cosh(s\omega_2) +3 }{2 \sinh (\omega_2s) }
\label{w1} 
\end{equation}
Note that the contribution from the common field always comes through the free 
particle term 
in the $P^2$ eigenvalue (\ref{E}) and so always gives rise to an overall factor .
The tachyonic instability \cite{bs},like that in the BFSS model \cite{lyf} is also present 
for very small separation.

We consider the potential for the above configuration already calculated from 
open string theory upto one loop. A 3-brane with electric field $F_1$ along 
$X^1$ and a magnetic field $F_2$ along $X^{23}$ plane interacts with a 1-brane
with the same electric field in its world volume through the term
$\int C \wedge tr(e^(iF))$ . 

The potential is given by \cite{lyf1}
\begin{equation}
A= \frac{L(1+F_1^2)}{2\pi}\int \frac{ds}{s} \frac{e^{-b^2s}}{4\pi s} B \times J ,
\end{equation}
where $B$ and $J$ are the contribution from the bosonic and the fermionic term
respectively and the prefactor takes care of the common electric field. Also 
we have taken $\alpha'=\frac{1}{2\pi}$.

$B$ and $J$ are given by
\begin{eqnarray}
B &=& \frac{1}{2} f_1^{-4}\Theta_1^{-1}(i\epsilon s)\nonumber\\ 
J &=& (
-f_2^4\Theta_2(i\epsilon s)  
+f_3^4\Theta_3(i\epsilon s)  
+f_4^4\Theta_4(i\epsilon s)  ) \label{A1}
\end{eqnarray}
where $\epsilon$ is related to the magnetic field by $\tan(\pi\epsilon) = F_2$

The matrix model describes the interaction through the open string and so let
us consider the short distance limit of this potential. 

If we write $\epsilon_i = \frac{\pi}{2} - \pi\omega_i$ where 
$\tan(\pi\epsilon_i) = F_i$
and consider the contribution of the massless modes only the in the limit
of large fields {\it i.e.} small $\omega$ the potential can be written as
\begin{equation}
W= L_1 \frac{1}{|2\pi \omega _1|^2}\int \frac{ ds}{s}
\exp(-b^2s)(\frac{ \pi}{2s}). \frac{
\cosh(2s\omega_2)-4\cosh(s\omega_2) +3 }{2 \sinh (\omega_2s) }
\label{w2} 
\end{equation}
which is  precisely the result obtained from the matrix model.

The comparison also makes it clear that, at least at the short distance limit,
the bosonic contribution comes from the $tre^{-sE}$ term while the $(-1)^F$ 
of the R sector contribution comes from the  
second term in the braces of (\ref{w5}). It is interesting that how the
latter will flip sign in the case of the brane-antibrane potential.

We can now calculate the long range potential from the matrix model
calculation for $p$, $q$-branes and compare with that obtained
in string theoretic calculation. A straightforward approximation
for the large value of $b$ in the expression for potential in (\ref{w})
leads to the expression
\begin{equation}
\frac{\pi}{2}^k\frac{\left [ \frac{1}{2} \sum _{i=k+1}^l \omega_i^4 - \sum_{i<j}
\omega_1^2\omega_j^2\right]}{2^{l-k}\prod_{i=k+1}^l \omega_i} \Gamma(4-l)b^{2l-8} \equiv V(b)
\label{lrv}
\end{equation}
Now we will see how this potential depends on b for the specific cases
. 

For 1-brane and 3-brane interaction, 
$ V(b) = \frac{\pi^4}{8} \frac{1}{F_{23}^3 b^4} $ 
which is a repulsive potential. Similarly for the case of
3-brane and 5-brane interaction ,
$ V(b) = \frac{\pi^4}{8} \frac{1}{F_{45}^3 b^4} $ .
Something interesting happens for the interaction between 1-brane and 5-brane.
In that case 
$V(b)= \frac{\pi(\omega_2^2-\omega_3^2)^2}{ 16 \omega_2\omega_3}$
$ V(b) = \frac{\pi^3}{b^4} \frac {(F_{23}^2 - F_{45}^2)} {F_{23}^3 F_{45}^3} 
$ .
and the potential vanishes for $F_{23} = F_{45}$. This is because when the
magnetic fields in the $X^{23}$ and $X^{34}$ planes are equal the 
configuration becomes a BPS state and so there is no force between them.

Other D-brane bound states can be analysed in a similar manner and those
also lead to a similar kind of agreement. Also one can consider the D-branes
oriented orthogonally to each other. Further all these results also
match with those obtained from BFSS model which is an evidence for the
duality between the matrix theories.

\section{ Scattering of branes}

In this section, we will consider the interaction between two moving branes of
 different dimensions. Again such a configuration is not a BPS state and
breaks the supersymmetry. We will calculate the phase shift of one brane 
(of lower dimension) while passing by the other in an eikonal approximation
\cite{lyfm,lyf} in the matrix model and then compare with the results
obtained by D-brane technique \cite{tasi,bac}.

Let us consider two branes of dimension $p$ and $q$ parallel to the $X^{1\cdots p}$
and $X^{1\cdots q}$ hyperplanes moving parallel to $X^{q+1}$ with a relative
velocity $v$ at a relative separation $b$ along $X^{q+2}$ axis. The classical
solutions corresponding to this configuration can easily be obtained by boosting 
the static solutions.   

These solutions are:

\begin{eqnarray}
A_0&=&\left ( \begin{array}{cc} 
B_0\cosh \epsilon  & 0\\ 0& B_0\cosh \epsilon \end{array}  \right );\ \ 
A_i =\left ( \begin{array}{cc} 
B_i&0\\ 0&B_i \end{array} \right ) ;\ \ {\mathrm for} \ \ 1\le i\le p 
 \nonumber \\
A_i &=& \left (\begin{array}{cc} 
B_i &0 \\0&0 \end{array} \right) , \ \ {\mathrm for} \ \ p<i\le q \ \ 
A_{q+1} = \left ( 
\begin{array}{cc} B_0\sinh \epsilon  &0\\0&B_0\sinh \epsilon  \end{array}
\right );\nonumber \\ 
A_{q+2} &=& \frac{ b}{2}
\left ( \begin{array}{cc} 1&0\\0&{-1} \end{array} \right ); \ \ 
A_i = 0 \mbox{ for } i\ge q+3.
\label{ }
\end{eqnarray}
where $B_i$'s have the usual meaning mentioned earlier,
$v \equiv \tanh \epsilon$ and $\epsilon$ is the boosting angle.

Now we consider the action of the solution on Hermitian matrices in the 
adjoint representation. This time the representations are more complicated 
than the previous one due to the presence of boost. The explicit forms
are
\begin{eqnarray}
P_0{\matrix X Y {Y^\dagger} Z} &=& 
i \omega _1 \cosh \epsilon  ( \partial
_1^1+\partial _1^2 ) {\matrix X Y {Y^ \dagger} Z} ;\ \ 
P_1 {\matrix X Y {Y^\dagger} Z} 
= (q_1^1 - q_1^2)
{\matrix X Y {Y^\dagger} Z} ;\ \ \nonumber \\ 
P_{q+1} {\matrix X Y {Y^\dagger} Z}
&=&i \omega _1\sinh \epsilon   \left ( \begin{array}{cc}(\partial _1^1+
\partial _1^2)X &( \partial _1^1- \partial ^2_1)Y\\
-(\partial _1^1-\partial _1^2 )Y^\dagger &-(\partial _1^1+\partial _1^2)
 Z \end{array} \right );\nonumber \\ 
 P_{p+1}
\left(\begin{array} {cc} X&Y\\ Y^\dagger &Z \end{array} \right) 
&=& i \omega _{k_1+1} \left ( \begin{array}{cc} ( \partial
^1_{k+1})+ \partial ^2_{k+1})X & \partial ^1_{k+1} Y \\
\partial^2_{k+1}Y^\dagger & 0 \end{array} \right
);\nonumber \\ 
 P_{p+2}
\left(\begin{array} {cc} X&Y\\ Y^\dagger &Z \end{array} \right) 
&=& \left ( \begin{array}{cc} (q_{k+1}^1-q^2_{k+1})X& q^1_{k+1}Y \\
-q^2_{k+1}Y^\dagger & 0 \end{array} \right );\ \   
P_{q+2}
\left(\begin{array} {cc} X&Y\\ Y^\dagger &Z \end{array} \right) 
= b
\left(\begin{array} {cl} 0&Y\\ Y^\dagger &Z \end{array} \right) 
;\nonumber \\
 P_i
\left(\begin{array} {cc} X&Y\\ Y^\dagger &Z \end{array} \right) 
&=&0; i\ge q+3 
\label{ }
\end{eqnarray}
where $q^1$ and $q^2$ are the first and the second arguments of the 
infinite dimensional matrices.

We evaluate the action of $F_{ij}$ in the adjoint representation and it has the 
non-zero action only on the matrices $Y$.
\begin{equation}
F_{1,q+1} 
\left(\begin{array} {cc} X&Y\\ Y^\dagger &Z \end{array} \right) 
 = 2 \omega _1 \sinh \epsilon  
\left(\begin{array} {cc} 0&Y\\ Y^\dagger &0 \end{array} \right) 
;\ \ F_{2i-2,2i} 
\left(\begin{array} {cc} X&Y\\ Y^\dagger &Z \end{array} \right) 
  =  -i \omega _i
\left(\begin{array} {cc} 0&Y\\ -Y^\dagger &0 \end{array} \right)
;\ \ i=k+1 \cdots q
\label{ }
\end{equation}

Therefore $F_{ij}$ gives nontrivial results only when it acts on $Y$ .
So we plug in the above expressions in the equation (\ref{w}) 
and get the phase shift to be
\begin{eqnarray}
Re W &=& \int_0^\infty \frac{ ds}{s} Tr e^{-sE} ( 
\cosh(2s\omega_1\sinh\epsilon)
+ \sum_{k+1}^l \cosh(2s\omega_i) \nonumber \\
&-& 2^{1-\Delta}2\cosh(2s\omega_1\sinh\epsilon)\prod_{k+1}^l 
2\cosh(s\omega_i) +3-\Delta )
\label{w }
\end{eqnarray}
where $\Delta=l-k$.

The eigen values of $P^2$ in the adjoint representation can be obtained
in a straightforward manner which again represents the Hamiltonian of some 
free particles and some harmonic oscillators.

\begin{equation}
E = 2p_1^2\cosh^2 \epsilon  + 2\sum_{i=2}^k (p_i^2+\tilde q_i^2) +
\sum_{i=k+1}^l[p_i^2+q_i^2] + 2[p_1^2\sinh^2 \epsilon  +q_i^2] + b^2
\end{equation}
where $[p_i, q_j] = -i\omega_i\delta_{ij}$ for $k<i,j\le l$ and all other
commutators vanish.

Now we calculate $Tr e^{-sE}$ as
\begin{equation}
Tr e^{-sE} = \frac{ N^{2k-1} L_1}{L_0 L_1\cdots L_p} \left ( \frac{
\pi}{2s} \right )^{\frac{ p}{2}}\frac{ 1}{\cosh \epsilon  }
\frac{ e^{-b^2s}}{2\sinh(s \omega _1\sinh\epsilon)
\prod_{i=k+1}^l 2\sinh(2s\omega_i)}
\label{tr2}
\end{equation}

Substituting this expression for Trace in \ref{w5} we get 
\begin{equation}
Re W = -V_p 2\pi \omega _i\prod_i^k \frac{ 1}{(2\pi \omega
_i)^2}\int_0^\infty \frac{ ds}{s} e^{-b^2s}( \frac{ \pi}{2s})^{\frac{
p}{2}}
\label{rew}
\end{equation}

So the expression of the phase shift can be written as
\begin{eqnarray}
&&Re W = -V_p 2^{\frac{-p}{2}} \omega _1 \prod_{i=1}^k \frac{ 1}{ \omega _i^2}
\int_0^\infty
\frac{ ds}{s} e^{-b^2 S} (4\pi s)^{\frac{ -p}{2}} \nonumber \\
&&\times \frac{ 
\cosh(2s\omega_1\sinh\epsilon)
+ \sum_{k+1}^l \cosh(2s\omega_i)
-2^{1-\Delta}2\cosh(s\omega_1\sinh\epsilon)\prod_{k+1}^l 
2\cosh(s\omega_i) +3-\Delta }
{\cosh \epsilon\  2\sinh(2s \omega _1\sinh\epsilon)
\prod_{i=k+1}^l 2\sinh(s\omega_i)}
\label{sv}
\end{eqnarray}
where $V_p$ represents the space time volume of the $p$-brane.

In order to compare with the open string result it is better to consider a specific
configuration. We consider a simple one -- in between a 3-brane and a 1-brane.
So we take $p=1$ and $q=3$. Also as in the last section the 3-brane has non-zero
magnetic field along the$X^{23}$plane and each of the branes has an electric
field in $X^1$ direction. 

Then the phase shift gets reduced to
\begin{eqnarray}
&&Re W = -\frac{V_1}{4\sqrt2}\omega _1 \frac{ 1}{ \omega _1^2}\int_0^\infty
\frac{ ds}{s} e^{-b^2 s} (4\pi s)^{\frac{ -1}{2}} \nonumber \\
&\times& \frac{ 2\cosh(2s\omega_1\sinh\epsilon)+ 2\cosh(2s\omega_2) - 
8\cosh (s\omega_1\sinh\epsilon)\cosh(s\omega_2) + 4}{
2\cosh \epsilon \sinh (2s \omega _1\sinh \epsilon)
2\sinh (s \omega _2)}
\end{eqnarray}

It is difficult to obtain the scattering phase shift between two
D-brane configurations with electric field in their world volume from the 
open string calculation. The presence of the electric field and the velocity
mixes up the coordinates and it is complicated to get a basis in which the 
energy momentum tensor can be diagonalized.
But we have seen, in the case of the potential and also in the cases of scattering
with a magnetic field, that the matrix model result gives the correct short
distance behavior. So we can demand that the matrix model gives the
phase shift for branes with electric fields in world volume, at least
at the limit of infinite field strength.

In the case of scattering of p, q-branes we can also  calculate the long
range potential from matrix model calculations.
This can be obtain from (\ref{sv})
by considering large $b$ approximation.
\begin{equation}
\frac{\tanh\epsilon\prod_{i=k+1}^l\frac{1}{a_i}\sum_{i=k}^la_i^4 - 
 2\sum_{i<j} a_i^2a_j^2}{a_k\prod a_i b^{7-q}} \equiv V^s(b)
\label{lrsp}
\end{equation}
where $a_i = \omega_i $ for $k+1\le i<l$ and $a_k = \omega_1\sinh\epsilon$.
This is the effective potential responsible for the scattering. In the case
of two branes of same dimension it gets reduced to
\begin{equation}
V^s(b,\epsilon) = \omega_1^2\sinh^2\epsilon\tanh^2\epsilon b^{q-7}
\end{equation}
In the low velocity limit the leading order term is 
$V(b,v)=\frac{\pi^2v^4}{F_{01}^2 b^{7-p}}$. So it reproduces the correct leading
order behavior of the potential. For electric 
field becomes infinite, the boundary conditions becomes Dirichlet and
so the potential vanishes. 

Now let us consider the specific cases:
For 1-brane and 3-brane scattering case , 
$$V^s(b,\epsilon) = \frac{\tanh\epsilon\left 
( \omega_1^2\sinh^2\epsilon -\omega_2^2\right)^2}
{\omega_1\omega_2^2\sinh\epsilon} b^{-4}.$$
Therefore  for $\sinh \epsilon = \frac{F_{01}}{F_{34}}$ again the potential
will vanish giving rise to BPS state.
In the low velocity limit, $V(b,v) \sim \frac{\pi F_{01}}{F_{34}^2 b^4}
(1-\frac{F_{34}^2}{F_{01}^2}v)^2$. 
Similar interpretation will hold good for the case of 3-brane scattering with
5-brane. 

For $p=1$, $q=5$ the potential that corresponds to the phase shift
is given by 
\begin{eqnarray}
V(\epsilon ,b) \sim \frac{\tanh\epsilon}{\omega^2_1 \sinh\epsilon}
\times \frac{1}{\omega_2^3\omega_3^2} 
 (\omega_1\sinh\epsilon - \omega_2 + \omega_3)
&&(\omega_1\sinh\epsilon + \omega_2 - \omega_3)\nonumber \\
&\times &(\omega_1\sinh\epsilon - \omega_2 - \omega_3)
(\omega_1\sinh\epsilon + \omega_2 + \omega_3)
\times b^{-2}\nonumber \\ 
\end{eqnarray}

So for $\sinh\epsilon = \pm\frac{\omega_2 - \omega_3}{\omega_1} ,
\pm\frac{\omega_2 + \omega_3}{\omega_1} $ the potential will vanish. 

In the low velocity limit it becomes
$$
V(b,v)=\frac{\omega_1}{(\omega_1\omega_2\omega^3)^2}
[\omega_1^2v^2-(\omega_2-\omega_3)^2]
[\omega_1^2v^2-(\omega_2+\omega_3)^2]
$$
It is interesting that here the potential changes its sign twice with the 
increase of the velocity.

The matrix model results, as mentioned earlier, is expected to
reproduce the short distance behavior. For these cases of bound states
of D-branes this should be true at least in the limit of the large
electromagnetic field which we have checked in the case of the
potential. On the basis of the presence of large field we can similarly
argue \cite{lyfm,lyf} that the long distance limit should also be reproduced
reliably. It will be interesting to check these scattering of the bound
states of D-branes (with electric field) in the closed string theory.
\section{Discussion}

In this paper, we have discussed the potential between D-branes with
electromagnetic fields turned on in its world volume in the matrix model
framework. This sort of configuration correspond to the stack of lower
dimensional branes attached to it and thus forming a non-threshold bound
state. Such a configuration, usually, does not saturate the BPS bound but
we have seen that they correspond to the classical solutions of the matrix
model which breaks half of the supersymmetry. The fact that the number of 
the lower dimensional branes is infinity makes this possible.

The interaction of these kind of objects have been discussed in \cite{lyf1}
from the open string calculation. We have seen that our result agrees 
with those obtained from open string calculation at the short distance 
limit. This is expected since the matrix model takes care of the massless
modes of the open strings and the open strings are known to dominate the
dynamics at the short distance limit. We have matched our result with the
long distance limit also which is dominated by the closed strings. This is
again due to the fact that the electromagnetic fluxes are very large.

Apart from getting agreement with the string theory, we have also noted the 
similarity between the potential between the odd and the even dimensional 
branes obtained from the two different matrix models which is an
evidence for the duality between the two matrix models. 
This is also expected since they
describe the IIA and IIB theories and the dynamics are also of the D-particles
and the D-instantons which are T-dual to each other.

We have also calculated the scattering phase shift between branes of odd
dimensions with electromagnetic fields. Due to the presence of the electric
fields {\it i.e} the D-instantons attached to the branes, this type of 
scattering is difficult to study in the open string calculation. On the other 
hand it is easier to study it in the matrix model framework. On the basis
of other agreements we can propose the result to be consistent with
the true short distance behavior at the limit of large field. The long
distance limit is also interesting and for a few choices of
velocities the potential obtained from the phase shift vanishes, signaling
existence of BPS saturation. A similar investigation in the closed
string theory may be interesting.

Finally, we have considered the trivial classical solution of the matrix model.
There are non-trivial solutions, such as instanton solutions which correspond
to other D-brane configurations and the the dynamics of these kind of objects 
can be studied in a similar manner. Also there are anti branes whose interaction
with the branes may be studied in this framework. However as mentioned earlier
in such a case the fermionic contribution should flip its sign and it is not
apparent how it will come about. 
By using the T-duality argument in the matrix model one can study
the interaction between the D-branes and the NS 5-branes of type IIB
theory in away similar to that used in \cite{lyf2}. Also it will be
interesting if the pure D-brane or a D-brane with finite flux turned on
can be constructed and studied in this matrix model framework.
  
{\bf Acknowlegement}:
It is a plesure to thank Dr. A. Kumar for useful discussions and for careful 
reading of the manuscript. We would also like to thank K. Roy for useful 
discussions and suggestions.

\end{document}